\documentclass[12pt]{article}
\usepackage{amssymb}
\usepackage{amsfonts}
\usepackage{amsmath}
\usepackage{titletoc}
\usepackage[abbr]{harvard}
\usepackage{setspace}
\usepackage{scalefnt}
\usepackage{graphicx}
\usepackage{tabularx,bm,booktabs,upgreek}
\newtheorem{theorem}{Theorem}

\newtheorem{proposition}[theorem]{Proposition}

\renewcommand{\cite}{\citeasnoun}
\allowdisplaybreaks

	 \small \normalsize%
\title{ \Large \textbf{Nonparametric Tests of Tail Behavior\\ in Stochastic Frontier Models} }

\author{William C. Horrace\thanks{Corresponding author: Department of Economics and Center for Policy Research, Syracuse University, Syracuse, NY, 13244, whorrace@syr.edu.} \qquad Yulong Wang\thanks{Department of Economics and Center for Policy Research, Syracuse University, Syracuse, NY, 13244.} }

\begin{document}
\maketitle
\vspace{10mm}
	\begin{abstract}
		This article studies tail behavior for the error components in the stochastic frontier model, where one component has bounded support on one side, and the other has unbounded support on both sides. Under weak assumptions on the error components, we derive nonparametric tests that the unbounded component distribution has thin tails and that the component tails are equivalent. The tests are useful diagnostic tools for stochastic frontier analysis. A simulation study and an application to a stochastic cost frontier for 6,100 US banks from 1998 to 2005 are provided. The new tests reject the normal or Laplace distributional assumptions, which are commonly imposed in the existing literature. \\
			
		
\end{abstract}
\newpage
	
	\doublespacing
	
\section{Introduction}
Stochastic frontier analysis (SFA) has a vast literature, both methodological and applied, and empiricists have applied the methods to myriad industries, most notably agriculture, banking, education, healthcare, and energy. A common practice in SFA is to impose parametric assumptions on the error components, but the set of statistical tools to investigate the validity of these assumptions is still limited. This paper expands this set of tools by drawing on recently developed techniques in Extreme Value (EV) theory and by developing new diagnostic tests.

In particular, the parametric stochastic frontier model for cross-sectional data (Aigner et al. 1997) is a leading case of the error component regression model but with the unique feature that one error component ($U$) is a non-negative random variable (e.g., half-normal, exponential), while the other ($W$) is a random variable of unbounded support (e.g., normal, Laplace, Student-t). A common assumption in the stochastic frontier literature is that $W$ is drawn from a normal or Laplace distribution (both thin-tailed distributions). See Aigner et al. (1977) or Horrace and Parmeter (2018), respectively.\footnote{For other parametric specifications of the model see Li (1996), Carree (2002), Tsionas (2007), Kumbhakar et al. (2013), and Almanidis et al. (2014). } However, heavy-tailed distributions are now also being considered. For example, the findings of Wheat et al. (2019) suggest that a cost inefficiency model of highway maintenance costs in England has Student-t errors.\footnote{There are semi-parametric estimators of the model that relax the distributional assumptions on one component and estimate the density of the other using kernel deconvolution techniques. See Kneip et al. (2015), Horrace and Parmeter (2011), Cai et al. (2020), Simar et al. 2017, Hall and Simar (2002), Florens et al. (2020).}  These parametric distributions, such as normal and Student-t, display similar patterns in the middle of their supports but exhibit substantially different tail behaviors. This observation motivates and plays an essential role in our diagnostic tests, which we believe are a timely and appropriate contribution to the literature.

The key idea of our test is as follows. Assuming independence of the error components, the largest order statistics of the composed error term ($Z = W-U$) (approximately) arise from the tails of $W$, because $U$ is one-sided. Also, assuming that $W$ is in the domain of attraction (DOA) of extreme value distributions, the asymptotic distribution of the largest order statistics of $W$ is the EV distribution, which may be fully characterized (after location and scale normalization) by a single parameter that captures its tail heaviness.\footnote{The assumption that $W$ is in the DOA of extreme value distributions is not restrictive, as we shall see.} Then, likelihood ratio statistics for hypotheses on this single parameter can be derived based on the limiting EV distribution.

To be specific, consider the right tail of $W$. If the DOA assumption is satisfied, then tail behavior may be entirely characterized by a \textit{tail index}, $\xi \in \mathbb{R}$. If $\xi = 0$, then $W$ has thin tails. If $\xi > 0$, then $W$ has thick tails. Otherwise, $W$ has bounded support. Under very weak assumptions on the error components, we derive a test that the tails of $W$ are thin ($\xi=0$). We prove that this test is valid whether $Z$ is observed or appended to a regression model (as it is in the stochastic frontier model). If we assume that $U$ is also in the DOA of extreme value distributions and that $W$ is symmetric (a common assumption), we also derive a test that the (right) tail of $U$ is thinner than the left tail of $W$. If we further assume that $W$ is a member of the normal family, then we may test the hypotheses that the tails of $U$ and $W$ are both thin. Therefore, our nonparametric tests are useful diagnostic tools to help empiricists make parametric choices on the distributions of both $U$ and $W$. This is particularly important for the stochastic frontier model for cross-sectional data, where distributional assumptions on the components are typically necessary for the identification of the model's parameters.

The paper is organized as follows. The next section presents the tests. Section 3 provides a simulation study of the power and size of the test. Section 4 applies the tests to a stochastic cost function for a panel of US banks 1998-2005, revealing that the tails of $W$ are not thin. Therefore, a normal or Laplace assumption for $W$ is not justified, and perhaps a Student-t assumption may be appropriate. Section 5 concludes.

\section{Tests of Tail Behavior}
To fix ideas, we begin a review of the DOA assumption and present the test in the case where $Z$ is directly observed in Section \ref{sec: no covariate}. Then in Section \ref{sec: covariate}, we move to the case where $Z$ is appended to a regression model and has to be estimated, which covers the linear regression stochastic frontier model. Additional tests under different sets of weak assumptions are also presented.\footnote{While the analyses that follow are for cross-sectional data, they can easily be applied to panel data, as long as one is willing to assume independence in both the time and cross-sectional dimensions.}
\subsection{The case with no covariates}\label{sec: no covariate}
Consider a random sample of $Z_{i}=W_{i}-U_{i}$ for $i=1,\ldots ,n$, where $U_{i}\geq 0$ represents \textit{inefficiency}, and $W_{i}\in \mathbb{R}$ is \textit{noise} with unbounded support. We start with testing the \textit{shape} of the right tail of $W_{i}$ in a nonparametric way.

The key assumption is that the distribution of $W_{i}$ is within the domain
of attraction of EV distributions. In particular, a cumulative distribution function $F$
is in the domain of attraction of $G_{\xi }$, denoted as $F\in \mathcal{D}%
\left( G_{\xi }\right) $, if there exist constants $a_{n}>0$ and $b_{n}$
such that 
\begin{equation*}
\lim_{n\rightarrow \infty }F^{n}\left( a_{n}v+b_{n}\right) =G_{\xi }\left(
v\right)
\end{equation*}%
where $G_{\xi }$ is the generalized EV distribution,
\begin{equation}
G_{\xi }(v)=\left\{ 
\begin{array}{l}
\exp [-(1+\xi v)^{-1/\xi }]\text{, }1+\xi v\geq 0\text{, for }\xi \neq 0 \\ 
\exp [-e^{-v}]\text{, }v\in 
%TCIMACRO{\U{211d} }%
%BeginExpansion
\mathbb{R}
%EndExpansion
\text{, }\xi =0%
\end{array}%
\right.  \label{def_G}
\end{equation}%
and $\xi $ is the tail index, measuring the decay rate of the tail.

The domain of attraction condition is satisfied by a large range of commonly
used distributions. If $\xi $ is positive, this condition is equivalent to
regularly varying at infinity, i.e., 
\begin{equation}
\lim_{t\rightarrow \infty }\frac{1-F\left( tv\right) }{1-F\left( t\right) }%
=v^{-1/\xi }\text{ \ \ \ \ for }v>0.  \label{RV}
\end{equation}%
This covers Pareto, Student-t\footnote{The tail index of the Student-t distribution with $\nu$ degrees of freedom is $\xi =1/\nu$.}, and F distributions, for example. The case
with $\xi =0$ covers the normal family, and the case with $\xi <0$
corresponds to distributions with a bounded support.\footnote{%
The uniform distribution has $\xi =-1$, and the triangular distribution has $%
\xi =-1/2$.} See de Haan and Ferreira (2007), Ch.1 for a complete review.

Note that the above notation is for the right tail of $W$, which can be easily adapted to the left tail by considering $-W$. For expositional simplicity, we denote $\xi_{W_{-}}$ and $\xi_{W_{+}}$ as the tail indices for the left and right tails of $W$, respectively. The same notation applies to other variables (e.g., $U$ and $Z$) introduced later.

Returning to SFA, a common assumption is that $W_{i}$ is normal
or Laplace, which implies that $\xi_{W_{+}} =0$. So our hypothesis testing problem
is as follows:%
\begin{equation} \label{eq: hypo1}
H_{0}:\xi_{W_{+}} =0\text{ against }H_{1}:\xi_{W_{+}} >0. 
\end{equation} %
If the null hypothesis is rejected, we would then argue that some heavy-tailed
distribution should be used to model the noise and maybe the inefficiency as
well.

To obtain a feasible test, we argue that, since $U_{i}$ is bounded from below at zero, the largest order statistics of $%
Z_{i} $ are approximately stemming from the right tail of $W_{i}$. This is
formalized in Proposition \ref{prop evt}, which requires the following conditions. Let $Z_{n:n}\geq ,\ldots ,\geq
Z_{n:1}$ be the order statistics of $\{Z_{i}\}_{i=1}^{n}$ by descend
sorting. Denote%
\begin{equation*}
\mathbf{Z}_{+}=\left( Z_{n:n},...,Z_{n:n-k+1}\right) ^{\intercal }
\end{equation*}%
as the $k$ largest observations. From now on, we use bold letters to denote vectors.
Denote $F_{W}$ and $Q_{W}(p)=\inf \{y\in 
%TCIMACRO{\U{211d} }%
%BeginExpansion
\mathbb{R}
%EndExpansion
\left. :\right. p\leq F_{W}(y)\}$ as the CDF and the quantile function of $%
W_{i}$, respectively. Write $Q_{W}(1)$ as the right end-point of the support
of $W_{i}$. For a generic column vector $\mathbf{X}$ and scalar $c$, the notation $%
\mathbf{X}-c $ means $\mathbf{X}-(c,\ldots ,c)^{\intercal }$.

\paragraph{Assumption 1}

\begin{description}
\item (i) $\left( U_{i},W_{i}\right) ^{\intercal }$ is \textit{i.i.d.}\ 

\item (ii) $U_{i}$ and $W_{i}$ are independent.

\item (iii) $U_{i}\geq 0$ with $\mathbb{E}\left[ \left\vert U_{i}\right\vert %
\right] <\infty $ and $W_{i}\in 
%TCIMACRO{\U{211d} }%
%BeginExpansion
\mathbb{R}
%EndExpansion
$ with $Q_{W}(1)=\infty $.

\item (iv) $F_{W}\in \mathcal{D}(G_{\xi_{W_{+}} })$ with $\xi_{W_{+}} \geq 0$. In addition, $%
F_{W}(\cdot )$ is twice continuously differentiable with bounded
derivatives, and the density $f_{W}(\cdot )$ satisfies that $\partial
f_{W}(t)/\partial t\nearrow 0$ as $t\rightarrow \infty $ on $[c,\infty )$
for some constant $c$.
\end{description}

Assumptions 1(i)-(iii) are common in the SFA literature (see Horrace and Parmeter (2018) and the references therein). Assumption 1(iv) requires the tail of $F_{W}$ to be within the domain of
attraction of EV distributions with an infinite upper bound. Moreover, it requires
that the density derivative monotonically increases to zero. This is a mild
assumption and is satisfied by many commonly used distributions. For example,
the normal distribution is covered as seen by%
\begin{equation*}
\frac{\partial f_{W}(t)}{\partial t}\propto -t\exp (-\frac{t^{2}}{2}%
)\nearrow 0\text{ as }t\rightarrow \infty ,
\end{equation*}%
and the Pareto distribution is covered as seen by%
\begin{equation*}
\frac{\partial f_{W}(t)}{\partial t}\propto (-\alpha -1)x^{-\alpha
-2}\nearrow 0\text{ as }t\rightarrow \infty \text{ for some }\alpha >0.
\end{equation*}

Under Assumption 1, the following proposition derives the asymptotic distribution of $\mathbf{Z}_{+}$.

\begin{proposition}
\label{prop evt}
Suppose Assumption 1 holds. Then, there exist sequences of
constants $a_{n}$ and $b_{n}$ such that for any fixed $k$%
\begin{equation*}
\frac{\mathbf{Z}_{+}-b_{n}}{a_{n}}\overset{d}{\rightarrow }\mathbf{V}%
_{+}=\left( V_{1},...,V_{k}\right) ^{\intercal }
\end{equation*}%
where the joint density of $\mathbf{V}_{+}$ is given by $f_{\mathbf{V}_{+}|\xi_{W_{+}}
}(v_{1},...,v_{k})=G_{\xi_{W_{+}} }(v_{k})\prod_{i=1}^{k}g_{\xi_{W_{+}} }(v_{i})/G_{\xi_{W_{+}}
}(v_{i})$ on $v_{k}\leq v_{k-1}\leq \ldots \leq v_{1}$, and $g_{\xi_{W_{+}}
}(v)=dG_{\xi_{W_{+}} }(v)/dv$.
\end{proposition}

The proof is in Appendix A. This proposition implies that the distributions of $Z_{i}$ and $W_{i}$ share
the same (right) tail shape, which is entirely characterized by the tail index $\xi_{W_{+}} $. Such tail equivalence does not hold, however, for the left tails due to the existence of $U$. This is studied in Section \ref{sec sym} under the additional assumption that $W$ is symmetric.

If the constants $a_{n}$ and $b_{n}$ were known, $\mathbf{Z_{+}}$ is then
approximately distributed as $\mathbf{V_{+}}$, and the limiting problem is
reduced to the well-defined finite sample problem: constructing some inference
method based on one draw $\mathbf{V_{+}}$ whose density $f_{\mathbf{V_{+}}|\xi_{W_{+}} }$ is
known up to $\xi_{W_{+}} $. However, $a_{n}$ and $b_{n}$ depend on $F_{W}$ and hence are unknown \textit{a priori}.

To avoid the need for knowledge of $a_{n}$ and $b_{n}$, we consider the following
self-normalized statistic 
\begin{eqnarray}
\mathbf{Z}_{+}^{\ast } &=&\frac{\mathbf{Z}_{+}-Z_{n:n-k+1}}{%
Z_{n:n}-Z_{n:n-k+1}}  \label{max_inv} \\
&=&\left( 1,\frac{Z_{n:n-1}-Z_{n:n-k+1}}{Z_{n:n}-Z_{n:n-k+1}},...,\frac{%
Z_{n:n-k+2}-Z_{n:n-k+1}}{Z_{n:n}-Z_{n:n-k+1}},0\right) ^{\intercal }.  \notag
\end{eqnarray}%
It is easy to establish that $\mathbf{Z}^{\ast }$ is maximally invariant with
respect to the group of location and scale transformations (cf., Lehmann and Romano (2005), Ch.6). In words, the
estimator constructed as a function of $\mathbf{Z}^{\ast }$ remains
unchanged if data are shifted and multiplied by any non-zero constant. This
makes senses since the tail shape should be preserved no matter how data are
linearly transformed. This invariance property allows us to construct nonparametric tests for a stochastic frontier model that is otherwise not identified without parametric assumptions on $U$ and $W$.\footnote{In particular the non-zero expectation of $U$ precludes identification of unknown parameter $\delta$ in the model $Z_i=\delta+W_i-U_i$. } As such, our tests do not reveal anything about the location or the scale of the error components.

The continuous mapping theorem and Proposition \ref{prop evt} imply that for
any fixed $k$, as $n\rightarrow \infty $,%
\begin{equation*}
\mathbf{Z}_{+}^{\ast }\overset{d}{\rightarrow }\mathbf{V}_{+}^{\ast }\equiv 
\frac{\mathbf{V}_{+}-V_{k}}{V_{1}-V_{k}}.
\end{equation*}%
The CDF\ of $\mathbf{V}_{+}^{\ast }$ can be calculated via change of
variables as%
\begin{equation}
f_{\mathbf{V}_{+}^{\ast }\mathbf{|}\xi_{W_{+}} }\left( \mathbf{v}_{+}^{\ast }\right) =\Gamma
\left( k\right) \int_{0}^{b_{0}\left( \xi_{W_{+}} \right) }t^{k-2}\exp \left(
-(1+1/\xi_{W_{+}} )\sum_{i=1}^{k}\log (1+\xi_{W_{+}} v_{i}^{\ast }t)\right) dt,
\label{density_xs}
\end{equation}%
where $\mathbf{v}_{+}^{\ast }=(v_{1}^{\ast },\ldots ,v_{k}^{\ast })$, $%
b_{0}\left( \xi \right) =\infty $ if $\xi \geq 0$ and $-1/\xi $ otherwise,
and $\Gamma \left( k\right) $ is the gamma function. Note that the
invariance restriction costs two degrees of freedom since the first and last
elements of $\mathbf{V}_{+}^{\ast }$ are always 1 and 0, respectively. We
calculate this density by numerical quadrature.

Given $f_{\mathbf{V}_{+}^{\ast }\mathbf{|}\xi_{W_{+}} }$, we can construct the
generalized likelihood-ratio test for problem (\ref{eq: hypo1}). Since the alternative hypothesis is composite, we follow Andrews and Ploberger (1994) and Elliott et al. (2015) to consider the weighted average alternative 
\begin{equation*}
\int f_{\mathbf{V}_{+}^{\ast}|\xi_{W_{+}}}(\cdot)w(\xi_{W_{+}} )d\xi_{W_{+}}, 
\end{equation*}
where $w(\cdot )$ is a weighting function that reflects the importance of
rejecting different alternative values. Then our test is constructed as \footnote{In later sections, we set $w(\cdot )$ to be the standard uniform distribution over $(0,1)$ for simplicity.} 
\begin{equation}
\varphi (\mathbf{v}_{+}^{\ast })=\mathbf{1}\left[ \frac{\int f_{\mathbf{V}_{+}^{\ast
}|\xi_{W_{+}} }(\mathbf{v}_{+}^{\ast })w(\xi_{W_{+}} )d\xi_{W_{+}} }{f_{\mathbf{V}_{+}^{\ast }|0}(\mathbf{v}_{+}%
^{\ast })}>\mathrm{cv}(k,\alpha )\right] ,  \label{test}
\end{equation}%
 where the critical value $\mathrm{cv}(k,\alpha )$ depends on $k$ and the level of significance $\alpha $. We can
obtain it by simulation. By Proposition \ref{prop evt} and the continuous mapping theorem, this test controls size asymptotically as $\lim_{n\rightarrow \infty }\varphi (\mathbf{Z}_{+}^{\ast })=\alpha$.

\begin{figure}
	\centering
	\includegraphics[width=1.00\textwidth]{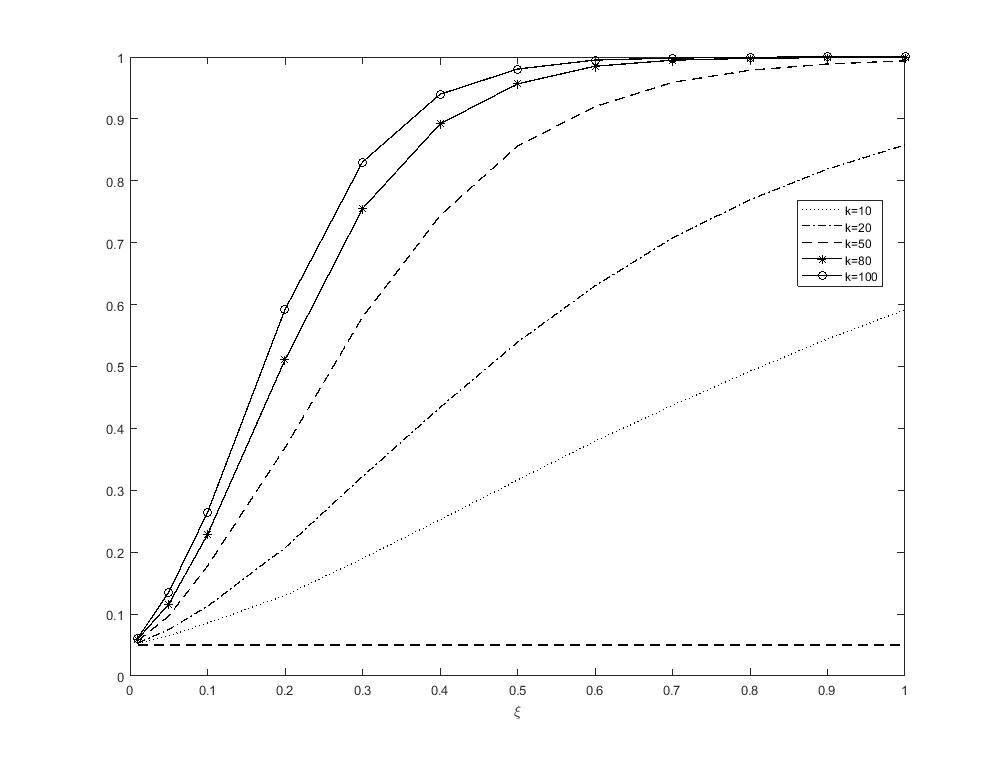}
	\caption{Asymptotic rejection probabilities of the test (\ref{test}) with $\mathbf{V}_{+}^{\ast }$ generated from the joint extreme value distribution (\ref{density_xs}) and the nominal size of $0.05$. The plots are based on numerical simulations with 10,000 random draws.}
	\label{fig:power_curves}
	${}$
\end{figure}

We end this subsection by briefly discussing the choice of $k$, that is, the number of the largest order statistics used to approximate the EV distribution. On the one hand, larger $k$ means including more mid-sample observations, which induces a larger finite sample bias in the EV approximation. On the other hand, smaller $k$ provides a better asymptotic approximation but uses less sample information, leading to a lower power test. This trade-off leads to difficulty in theoretical justification of an optimal $k$ in standard EV theory literature (cf., M\"{u}ller and Wang (2017)). It is even more difficult, if at all possible, in our case, since we only observe $Z$, and not $W$. Nonetheless, our asymptotic arguments show that the test (\ref{test}) controls size for any fixed $k$, as long as $n$ is sufficiently large. Figure \ref{fig:power_curves} depicts the asymptotic power of the test (\ref{test}) with $\mathbf{V}_{+}^{\ast}$ generated from the density (\ref{density_xs}) based on 10,000 simulation draws. The test controls size for all values of $k$ by construction and has reasonably large power when $k$ exceeds 20. 

With ideas fixed, we now turn to the regression version of the test, with application to SFA.

\subsection{The case with covariates: SFA}\label{sec: covariate}

Now consider the linear regression with 
\begin{equation*}
Y_{i}=\mathbf{X}_{i}^{\intercal }\beta _{0}+Z_{i},
\end{equation*}%
where $Z_{i}=-U_{i}+W_{i}$ is as in the previous section, and $\beta _{0}$ is
some pseudo-true parameter in some compact parameter space. This could be a Cobb-Douglas production function (in logarithms), where $Y$ is productive output and $U$ is now called \textit{technical efficiency}, which measures distance ($U_i$) from a stochastic frontier ($\mathbf{X}_{i}^{\intercal }\beta _{0}+W_{i}$). The slopes ($\beta_0$) are marginal products of the productive inputs, $\mathbf{X}_{i}$. It could also be a stochastic cost function if we multiply $U$ by $-1$. Suppose we have some estimator, $\hat\beta$ of $\beta_0$. The following assumption is imposed to construct our diagnostic test.

\paragraph{Assumption 2}

\begin{description}
\item (i) $\left( X_{i},U_{i},W_{i}\right) ^{\intercal }$ is i.i.d.\ 

\item (ii) $U_{i}$ and $W_{i}$ are independent.

\item (iii) $U_{i}\geq 0$ with $\mathbb{E}\left[ \left\vert U_{i}\right\vert %
\right] <\infty $ and $W_{i}\in \mathbb{R}$ with $Q_{W}(1)=\infty $.

\item (iv) $F_{W}\in \mathcal{D}(G_{\xi_{W_{+}} })$ with $\xi_{W_{+}} \geq 0$. In addition, $%
F_{W}(\cdot )$ is twice continuously differentiable with bounded
derivatives, and the density $f_{W}(\cdot )$ satisfies that $\partial
f_{W}(t)/\partial t\nearrow 0$ as $t\rightarrow \infty $ on $[c,\infty )$
for some constant $c$.

\item (v) $\left\vert \left\vert \hat{\beta}-\beta _{0}\right\vert
\right\vert \sup_{i}\left\vert \left\vert \mathbf{X}_{i}\right\vert
\right\vert =o_{p}(n^{\xi_{W_{+}} })$, if  $\xi_{W_{+}} >0$. $\left\vert
\left\vert \hat{\beta}-\beta _{0}\right\vert \right\vert \sup_{i}\left\vert
\left\vert \mathbf{X}_{i}\right\vert \right\vert /f_{W}\left(
Q_{W}(1-1/n)\right) \left. =\right. o_{p}(1)$, otherwise.
\end{description}

Assumption 2 is similar to Assumption 1 with additional restrictions on the covariate $\mathbf{X}$. In particular, Assumption 2(v) bounds the norm of $\hat{\beta}$ and $\left\vert \left\vert 
\mathbf{X}_{i}\right\vert \right\vert $. A sufficient condition when $\xi_{W_{+}} $
is positive is that $\left\vert \left\vert \hat{\beta}-\beta _{0}\right\vert
\right\vert =O_{p}(n^{-1/2})$ and $\sup_{i}\left\vert \left\vert \mathbf{X}%
_{i}\right\vert \right\vert =o_{p}(n^{1/2})$, which is easily satisfied in
many applications.\footnote{Even though $\mathbb{E}\left[ \left\vert U_{i}\right\vert \right] \ne 0$, ordinary least squares (OLS) will typically suffice for $\hat\beta$, because our test is invariant to relocation. } When $\xi $ is zero, we need slightly stronger bounds.
Straightforward calculations show that the normal distribution satisfies
Assumption 2(v) for the $\xi_{W_{+}} = 0$ case, if $\left\vert \left\vert \hat{\beta}-\beta
_{0}\right\vert \right\vert =O_{p}\left( n^{-1/2}\right) $ and $%
\sup_{i}\left\vert \left\vert \mathbf{X}_{i}\right\vert \right\vert \left.
=\right. O_{p}(n^{1/2-\varepsilon })$ for some $\varepsilon >0$. This is
seen by $1/f_{W}\left( Q_{W}\left( 1\left. -\right. 1/n\right) \right)
\left. \leq \right. O(\log (n))$ (cf.\ Example 1.1.7 in de Haan and Ferreira (2007)).

Denote $\hat{Z}_i$ as the OLS residuals and 
\begin{equation*}
\mathbf{\hat{Z}}_{+}=\left( \hat{Z}_{n:n},...,\hat{Z}_{n:n-k+1}\right) ^{\intercal }
\end{equation*} the largest $k$ order statistics. Then given Assumption 2, the following proposition derives the asymptotic distribution of $\mathbf{\hat{Z}}_{+}$
\begin{proposition}
\label{prop evtx}
Suppose Assumption 2 holds. Then, there exist sequences of
constants $a_{n}$ and $b_{n}$ such that for any fixed $k$%
\begin{equation*}
\frac{\mathbf{\hat{Z}}_{+}-b_{n}}{a_{n}}\overset{d}{\rightarrow }\mathbf{V}%
_{+}
\end{equation*}%
where the joint density of $\mathbf{V}_{+}$ is the same as in Proposition %
\ref{prop evt}.
\end{proposition}

The proof is in Appendix A. Proposition \ref{prop evtx} implies that the largest order statistics of the
regression residuals satisfy the same convergence as the no-covariate case. In other words, the estimation error from the OLS becomes negligible so that the largest order statistics are stemming from the right tail of $W$ asymptotically.
This validates the construction of the test (\ref{test}) by replacing 
$\mathbf{Z}_{+}^{\ast }$ with $\mathbf{\hat{Z}}_{+}^{\ast }$, where%
\begin{equation*}
\mathbf{\hat{Z}}_{+}^{\ast }=\frac{\mathbf{\hat{Z}}_{+}-\hat{Z}_{n:n-k+1}}{%
\hat{Z}_{n:n}-\hat{Z}_{n:n-k+1}}.
\end{equation*}
Proposition \ref{prop evtx} and the continuous mapping theorem, we similarly have  $\lim_{n\rightarrow \infty }\varphi (\mathbf{\hat{Z}_{+}}^{\ast })=\alpha$.

\subsection{Symmetry of noise $W$}\label{sec sym}

The previous analysis studies the right tail of $W$ (and equivalently $Z$). Suppose we assume $W$ has a symmetric distribution, then the tail indices of
both tails of $W$ become equivalent, and hence we can learn about the tail of $U$ using the left tail index
of $Z$. To this end, we make the following
additional assumption.

\paragraph*{Assumption 3}

\begin{description}
\item (i) $W_{i}$ is symmetric at zero.

\item (ii) $F_{U}\in \mathcal{D}( G_{\xi _{U_{+}}})$ with $\xi_{U_{+}} \geq0$.
\end{description}

Assumption 3(i) implies that $\xi_{W_{-}}=\xi_{W_{+}}$, and the condition that $U>0$ implies its left tail index is negative. Therefore, in this subsection only, we simply denote $\xi_{U}$ and $\xi_{W}$ as the right tail indices of $U$ and $W$, respectively. Now we can test if $U$ has a thinner or equal right tail than $W$ by specifying
the following hypothesis testing problem,%
\begin{equation}
H_{0}:\xi _{U}\leq \xi _{W}\text{ against }H_{1}:\xi _{U}>\xi _{W}\text{.}
\label{hypo uw}
\end{equation}%
Moreover, if $W$ is in the normal or Laplace family ($\xi _{W}=0$), since we limit the tail indices to be non-negative, the null hypothesis
then reduces to $\xi _{U}=\xi _{W}=0$.

Under the null hypothesis of (\ref{hypo uw}), $W$ is the leading term in $Z$
in both the left and right tails. Then the DOA
assumption for both $W$ and $U$ implies that $\xi _{Z_{-}}=\max \{\xi
_{U},\xi _{W}\}$, and Proposition \ref{prop evtx} entails $\xi
_{Z_{+}}=\xi _{W}$. Therefore, the above testing problem becomes equivalent to 
\begin{equation}
H_{0}:\xi _{Z_{-}}=\xi _{Z_{+}}\text{ against }H_{1}:\xi _{Z_{-}}>\xi
_{Z_{+}}\text{.}  \label{hypo equaltail}
\end{equation}

We now construct a test for (\ref{hypo equaltail}). Define $\mathbf{\hat{Z}}%
_{-}$ as the smallest $k$ order statistics of the estimation residuals, that is,
\begin{equation*}
\mathbf{\hat{Z}}_{-}=\left( \hat{Z}_{n:1},\hat{Z}_{n,2},\ldots ,\hat{Z}%
_{n,k}\right) ^{\intercal }
\end{equation*}%
and its self-normalized analogue as%
\begin{equation*}
\mathbf{\hat{Z}}_{-}^{\ast }=\frac{\mathbf{\hat{Z}}_{-}-\hat{Z}_{n:k}}{\hat{Z%
}_{n:1}-\hat{Z}_{n:k}}.
\end{equation*}%
The following proposition establishes that $\mathbf{\hat{Z}}_{-}^{\ast }$
asymptotically has the EV distribution with tail index $\xi _{Z_{-}}$ and is
independent from $\mathbf{\hat{Z}}_{+}^{\ast }$.

\begin{proposition}
\label{prop sym}Suppose Assumptions 2 and 3 hold. Then, for any fixed $k$,%
\begin{equation*}
\binom{\mathbf{\hat{Z}}_{-}^{\ast }}{\mathbf{\hat{Z}}_{+}^{\ast }}\overset{d}%
{\rightarrow }\binom{\mathbf{V}_{-}^{\ast }}{\mathbf{V}_{+}^{\ast }}\text{
as }n\rightarrow \infty ,
\end{equation*}%
where $\mathbf{V}_{-}^{\ast }$ and $\mathbf{V}_{+}^{\ast }$ are independent
and both EV distributed with density (\ref{density_xs}) and tail indices $\xi _{Z_{-}}$ and $\xi _{Z_{+}}$, respectively.
\end{proposition}

The proof is in Appendix A. Given the above proposition, we aim to construct a generalized likelihood ratio
test for (\ref{hypo equaltail}) as follows,
\begin{equation}
\varphi_{\pm} \left( \mathbf{v}_{-}^{\ast },\mathbf{v}_{+}^{\ast }\right) =%
\mathbf{1}\left[ \frac{\int_{\{\left( \xi _{-},\xi _{+}\right) \in \Xi
^{2}:\xi _{+}<\xi _{-}\}}f_{\mathbf{V}_{-}^{\ast }|\xi _{-}}(\mathbf{v}%
_{-}^{\ast })f_{\mathbf{V}_{+}^{\ast }|\xi _{+}}\left( \mathbf{v}_{+}^{\ast
}\right) w\left( \xi _{-},\xi _{+}\right) d\xi_{-}d\xi_{+}} {\int_{\Xi }f_{\mathbf{V}%
_{-}^{\ast }|\xi }(\mathbf{v}_{-}^{\ast })f_{\mathbf{V}_{+}^{\ast }|\xi
}\left( \mathbf{v}_{+}^{\ast }\right) d\Lambda \left( \xi \right) }>\mathrm{%
cv}(k,\alpha )\right] ,  \label{test eq}
\end{equation}%
where $\Xi$ denotes the parameter space of the tail indices, and $w\left( \cdot ,\cdot \right) $ is the weighting function for the
alternative hypothesis as in (\ref{test}). We set $\Xi$ to be $[0,1)$ to cover all distributions with a finite mean and $w(\cdot)$ to be uniform over the alternative space. The weight $\Lambda \left( \cdot \right) $ can be considered as the least favorable distribution, which we discuss more now.

Note that the null hypothesis of (\ref{hypo equaltail}) is composite. We need to
control size uniformly over all $\xi _{Z_{-}}=\xi _{Z_{+}}\in \Xi $. To that end, we can transform the composite null into a simple one by considering the weighted average density with respect to the
weight $\Lambda $. Together with a suitably chosen the critical value, this test (\ref{test eq}) maintains the uniform size control. Now the problem reduces to determining an appropriate weight $\Lambda$. Elliott et al. (2015) study the generic
hypothesis testing problem where a nuisance parameter exists in the null
hypothesis. We tailor their argument for our test (\ref{test eq}) and adopt
their computational algorithm for implementation. In particular, $\Lambda
\left( \cdot \right) $ and $\mathrm{cv}(k,\alpha )$ are numerically
calculated only once by the authors instead of the empiricists who use our test. They only need to construct the order statistics $\mathbf{\hat{Z}}_{-}^{\ast }$ and $\mathbf{\hat{Z}}_{+}^{\ast }$ and numerically evaluate the density. We
provide more computational details in the Appendix and the corresponding MATLAB code in the supplemental materials. By the
continuous mapping theorem and Proposition \ref{prop sym}, for any fixed $k$%
, $\lim \sup_{n\rightarrow \infty }\mathbb{E}\left[ \varphi _{\pm }\left( 
\mathbf{\hat{Z}}_{-}^{\ast },\mathbf{\hat{Z}}_{+}^{\ast }\right) \right]
\leq \alpha $ under the null hypothesis of (\ref{hypo equaltail}).

As we discussed above, the hypothesis testing problem (\ref{hypo equaltail})
simplifies to 
\begin{equation*}
H_{0}:\xi _{Z_{-}}=\xi _{Z_{+}}=0\text{ against }H_{1}:\xi _{Z_{-}}>\xi
_{Z_{+}}=0\text{,}
\end{equation*}%
if $W$ is assumed to be in the normal family ($\xi_{W}=0$). Proposition \ref{prop sym}
implies $\mathbf{\hat{Z}}_{-}^{\ast }$ and $\mathbf{\hat{Z}}_{+}^{\ast }$
are asymptotically independent and both of them are EV distributed. Then accordingly, our test (\ref{test eq}%
) reduces to%
\begin{equation*}
\mathbf{1}\left[ \frac{\int_{\Xi }f_{\mathbf{V}_{-}^{\ast }|\xi _{-}}(%
\mathbf{v}_{-}^{\ast })dW\left( \xi _{-}\right) }{f_{\mathbf{V}_{-}^{\ast
}|0}(\mathbf{v}_{-}^{\ast })}>\mathrm{cv}(k,\alpha )\right] ,
\end{equation*}%
which is identical to (\ref{test}). This suggests that we can simply substitute $%
\mathbf{\hat{Z}}_{-}^{\ast }$ into (\ref{test}) for implementation.

\section{Simulation Study}

\subsection{Hypothesis testing about noise $W$}

We set $w(\cdot )$ to be the uniform weight on $\left[ 0,0.99\right] $ to
include all distributions with a finite mean and the level of significance
to be $0.05$. In Table 1, we report the small sample rejection probabilities
of the test (\ref{test}). We generate $U_{i}$ from the right half-standard
normal and the right half-Laplace(0,1) distributions and $W_{i}$ from four
distributions: standard normal, Laplace(0,1) (denoted La(0,1)) Student-t(2),
Pareto(0.5) and F(4,4). The normal and Laplace distributions correspond to
the null hypothesis, and the other three are alternative hypotheses. The
results suggest that the test (\ref{test}) has an excellent performance in size
and power. Note that when $k=50$ and $n=100$, we essentially include too
many mid-sample observations so that the EV approximation is
poor.

Now we consider the linear regression model that $Y_{i}=\mathbf{X}%
_{i}^{\intercal }\beta _{0}+Z_{i}$ with $\mathbf{X}_{i}=\left(
1,X_{2i}\right) ^{\intercal }$ and $\beta _{0}=\left( 1,1\right) ^{\intercal
}$. We assume $X_{2i}\sim \mathcal{N}\left( 0,1\right) $ and independent
from $Z_{i}$. Table 2 reports the rejection probabilities of our test (\ref%
{test}). Findings are similar to those in Table 1.

\subsection{Hypothesis testing about inefficiency $U$ and noise $W$}

Consider the hypothesis testing problem (\ref{hypo equaltail}). We implement
the test (\ref{test eq}) with the same setup as above. Table 3 reports the
rejection probabilities under the null and alternative hypotheses. We
make the following observations. First, the test controls size well
unless $k$ is too large relative to $n$, as seen in the column with $n=100$
and $k=50$. This is again because we are using too many mid-sample observations to
approximate the tail so that the EV convergence in Propositions 1-3 provides
poor approximations. Second, the test has good power properties as seen from
the last five rows. In particular, using only the largest 50 order
statistics from 1000 observations leads to the power of 0.94. Finally, the
power decreases as the alternative hypothesis becomes closer to the null, as
we move down along rows.

Now we consider the special case where $W$ is in the normal family. Then
we implement (\ref{test}) with $\mathbf{\hat{Z}}_{-}^{\ast }$ as the input.
Table 4 contains the rejection probabilities under the null and alternative
hypotheses. The rows with $F_{U}$ being half-normal or Laplace correspond to
the size under the null hypothesis, while other rows the power under the
alternative hypothesis. The new test has excellent size and power properties.

\section{Empirical illustration}

We illustrate the new method using the US bank data collected by Feng and Serletis (2009). The data are a sample of US banks covering the period from 1998 to 2005 (inclusive). After deleting banks with negative or zero input prices, we are left with a balanced panel of 6,010 banks observed annually over the 8-year period. A more detailed description of
the data may be found in Feng and Serletis (2009). Here we specify a stochastic cost function, letting $Z=W+U$, so $U \geq 0$ is cost inefficiency, and more inefficient banks have higher total costs, $Y$. Since our tests are designed for cross-sectional data, we divide the original panel data into
cross-sections (one for each year) and regress the logarithm of total bank cost on
a constant and the logarithms of six control variables, including the wage
rate for labor, the interest rate for borrowed funds, the price of physical
capital, and the amounts of consumer loans, non-consumer loans, and
securities. Since the object of interest is the cost function, we multiply the OLS residuals by $-1$ and take the smallest and the largest $k\in\{25, 50, 75, 100\}$ order
statistics, respectively, to implement the test (\ref{test}). The p-values
are reported in Table 5. Under the assumption that $W$ is symmetric\footnote{The symmetry assumption is reasonable here and is imposed in Feng and Serletis (2009).}, these small p-values suggest that $W$ has heavy tails on both sides, so a Student-t assumption (e.g., Wheat, Stead, and Greene, 2019) is more appropriate.

\section{Concluding remarks}
We derive several nonparametric tests of the tail behavior of the error components in the stochastic frontier model. The tests are easy to implement in MATLAB and are useful diagnostic tools for empiricists. 

Often a first-step diagnostic tool for SFA is to calculate the skewness of the OLS residuals to see if they are properly skewed. See Waldman (1982), Simar and Wilson (2010), and Horrace and Wright (2020). If they are positively skewed, the maximum likelihood estimator of the variance of inefficiency is zero, and OLS is the maximum likelihood estimator of $\beta_0$. If they are negatively skewed, then OLS is not a stationary point in the parameter space of the likelihood, and the stochastic frontier model is well-posed. After calculating negatively skewed OLS residuals, a useful second-step diagnostic tool is to implement our nonparametric tests to understand the tail behaviors of the error component distributions and to guide parametric choices subsequently .

\newpage
\appendix

\section*{Appendix}
\section{Proofs}

\subsubsection*{Proof of Proposition \protect\ref{prop evt}}

Since only the right tail index of $W$ shows up in this proof, we simply denote $\xi = \xi_{W_{+}}$ in this proof.

We prove the case with $k=1$ first. By Corollary 1.2.4 and Remark 1.2.7 in 
de Haan and Ferreira (2007), the constants $a_{n}$ and $b_{n}$ can be chosen as follows.
If $\xi \left. >\right. 0$, we choose $a_{n}\left. =\right. Q_{W}(1\left.
-\right. 1/n)$ and $b_{n}(\xi )\left. =\right. 0$. If $\xi \left. =\right. 0$%
, we choose $a_{n}\left. =\right. 1/(nf_{W}(b_{n}))$ and $b_{n}\left.
=\right. Q_{W}(1\left. -\right. 1/n)$. By construction, these constants
satisfy that $1\left. -\right. F_{W}(a_{n}v\left. +\right. b_{n})\left.
=\right. O(n^{-1})$ for any fixed $v\left. >\right. 0$ in both cases (e.g., de Haan and Ferreira (2007), Ch.1.1.2).

By Assumption 1-(iv), we have that%
\begin{eqnarray*}
&=&\mathbb{P}\left( Z_{n:n}\leq a_{n}v+b_{n}\right) \\
&=&\mathbb{P}\left( Z_{i}\leq a_{n}v+b_{n}\right) ^{n} \\
&\equiv &A_{n}(v)\cdot \left( 1+\frac{B_{n}\left( v\right) }{\mathbb{P}%
\left( W_{i}\leq a_{n}v+b_{n}\right) }\right) ^{n},
\end{eqnarray*}%
where $A_{n}=\mathbb{P}\left( W_{i}\leq a_{n}v+b_{n}\right) ^{n}$, and%
\begin{equation*}
B_{n}(v)=\mathbb{P}\left( -U_{i}+W_{i}\leq a_{n}v+b_{n}\right) -\mathbb{P}%
\left( W_{i}\leq a_{n}v+b_{n}\right) \text{.}
\end{equation*}%
By Assumption 1-(iv), $A_{n}(v)\rightarrow G_{\xi }(v)$ for any constant $%
v>0 $. Then by the facts that $\mathbb{P}\left( W_{i}\leq
a_{n}v+b_{n}\right) \rightarrow 1$ and $(1+t/n)^{n}\rightarrow \exp (t)$, it
suffices to show that $B_{n}(v)=o\left( n^{-1}\right) $. To this end, we
have 
\begin{align*}
& \text{ \ \ }B_{n}(v) \\
& \left. =_{(1)}\right. \mathbb{E}\left[
F_{W}(a_{n}v+b_{n}+U_{i})-F_{W}(a_{n}v+b_{n})\right] \\
& \left. \leq _{(2)}\right. \sup_{t\in \left[ a_{n}v+b_{n},\infty \right]
}f_{W}(t)\cdot \mathbb{E}\left[ |U_{i}|\right] \\
& \left. \leq _{(3)}\right. f_{W}(a_{n}v+b_{n})\cdot \mathbb{E}\left[ |U_{i}|%
\right] \\
& \left. =_{(4)}\right. o(n^{-1}),
\end{align*}%
where eq.(1) is by Assumption 1-(ii) ($U_{i}$ is independent from $V_{i}$),
ineq.(2) is by the intermediate value theorem, ineq.(3) follows from
Assumption 1-(iv) ($f_{W}(t)$ is non-increasing when $t>c$ for some constant 
$c$), and eq.(4) is seen by Assumption 1-(iii) ($\mathbb{E}\left[ |U_{i}|\right]
<\infty )$ and Assumption 1-(iv). In particular, the fact that $%
nf_{W}(a_{n}v+b_{n})=o(1)$ is implied by the von Mises' condition. See, for
example, Corollary 1.1.10 in de Haan and Ferreira (2007) with $t=Q_{W}(1-1/n)$.

Generalization to $k>1$ is as follows. Consider $v_{1}>v_{2}>\cdots >v_{k}$.
Chapter 8.4 in Arnold et al. (1992) (p.219) gives that%
\begin{eqnarray*}
&&\mathbb{P}\left( Z_{n:n}\leq a_{n}v_{1}+b_{n},...,Z_{n:n-k+1}\leq
a_{n}v_{k}+b_{n}\right) \\
&=&F_{Z}^{n-k}(a_{n}v_{k}+b_{n})\prod_{r=1}^{k}\left( n-r+1\right)
a_{n}f_{Z}\left( a_{n}v_{r}+b_{n}\right) 
\\
&=&\left[ F_{W}^{n-k}\left( a_{n}v_{k}+b_{n}\right) \prod_{r=1}^{k}\left(
n-r+1\right) a_{n}f_{W}\left( a_{n}v_{r}+b_{n}\right) \right] \times \\
&&\left[ \left( \frac{F_{Z}\left( a_{n}v_{k}+b_{n}\right) }{F_{W}\left(
a_{n}y_{k}+b_{n}\right) }\right) ^{n-k}\prod_{r=1}^{k}\frac{f_{Z}\left(
a_{n}v_{r}+b_{n}\right) }{f_{W}\left( a_{n}v_{r}+b_{n}\right) }\right] \\
&\equiv &\tilde{A}_{n}\times \tilde{B}_{n}.
\end{eqnarray*}%
The convergence that $\tilde{A}_{n}\rightarrow G_{\xi }\left( v_{k}\right)
\prod_{r=1}^{k}\{g_{\xi }\left( v_{r}\right) /G_{\xi }\left( v_{k}\right)
\} $ is established by Theorem 8.4.2 in Arnold et al. (1992). It now remains to
show $\tilde{B}_{n}\rightarrow 1$. First, the fact that 
\begin{equation*}
(F_{Z}\left(
a_{n}v_{k}+b_{n}\right) /F_{W}\left( a_{n}v_{k}+b_{n}\right)
)^{n-k}\rightarrow 1
\end{equation*} 
is shown by the same argument as above in the $k=1$
case. Second, for any $v$%
\begin{eqnarray*}
\frac{f_{Z}\left( v\right) }{f_{W}\left( v\right) } &=&\frac{\frac{\partial 
\mathbb{E}\left[ F_{W}(v+U_{i})\right] }{\partial v}}{f_{W}\left( v\right) }
\\
&=&\frac{\frac{\partial }{\partial v}\int F_{W}(v+u)f_{U}\left( u\right) du}{%
f_{W}\left( v\right) } \\
&=&\frac{\int \frac{\partial }{\partial v}F_{W}(v+u)f_{U}\left( u\right) du}{%
f_{W}\left( v\right) }\text{ \ (by Leibniz's rule)} \\
&=&\frac{\mathbb{E}\left[ f_{W}\left( v+U_{i}\right) \right] }{f_{W}\left(
v\right) },\text{ }
\end{eqnarray*}%
where applying Leibniz's rule is permitted by Assumption 1-(iv), which
implies that $f_{W}\left( v\right) $ is uniformly continuous in $v$. Then
similarly as bounding $B_{n}$ above, we use the mean value expansion and
Assumptions 1(ii)-(iv) to derive that for any $r\in \{1,...,k\}$  and some
constant $0\left. <\right. C\left. <\right. \infty $, 
\begin{eqnarray*}
&&\left\vert \frac{f_{Z}\left( a_{n}v_{r}+b_{n}\right) }{%
f_{W}\left( a_{n}v_{r}+b_{n}\right) }-1\right\vert \\
&=&\left\vert \frac{\mathbb{E}\left[ f_{W}\left(
a_{n}v_{r}+b_{n}+U_{i}\right) -f_{W}\left( a_{n}v_{r}+b_{n}\right) \right] }{%
f_{W}\left( a_{n}v_{r}+b_{n}\right) }\right\vert \\
&\leq &\sup_{t\in \left[ a_{n}v_{r}+b_{n},\infty \right] }\left\vert \frac{%
\partial f_{W}\left( t\right) /\partial t}{f_{W}\left(
a_{n}v_{r}+b_{n}\right) }\right\vert \mathbb{E}\left[ |U_{i}|\right] \\
&\leq &\left\vert \frac{\partial f_{W}\left( a_{n}v_{r}+b_{n}\right)
/\partial t}{f_{W}\left( a_{n}v_{r}+b_{n}\right) }\right\vert \mathbb{E}%
\left[|U_{i}|\right] \text{ \ (by}\frac{\text{ }\partial f_{W}(t)}{\partial t}%
\nearrow 0\text{)} \\
&\leq &C\left\vert \frac{f_{W}\left( a_{n}v_{r}+b_{n}\right) }{1-F_{W}\left(
a_{n}v_{r}+b_{n}\right) }\right\vert \mathbb{E}\left[ |U_{i}|\right] \\
&=&o(1),
\end{eqnarray*}%
where the last inequality follows from the fact that $\lim_{t\rightarrow
\infty }\frac{\partial f_{W}\left( t\right) /\partial t(1-F_{W}(t))}{%
f_{W}\left( t\right) ^{2}}\rightarrow -1-\xi $, which is implied by the von
Mises's condition (cf. Theorem 1.1.8 in de Haan and Ferreira (2007)), and the last
equality follows from the facts that $n(1-F_{W}\left(
a_{n}v_{r}+b_{n}\right) )=O(1)$ and $nf_{W}\left( a_{n}v_{r}+b_{n}\right)
=o(1)$ (see again Corollary 1.1.10 in de Haan and Ferreira (2007) with $t=Q_{W}(1-1/n)$%
). The proof is then complete. $\blacksquare $

\subsubsection*{Proof of Proposition \protect\ref{prop evtx}}

In this proof, we drop the subscript $W_{+}$ in $\xi_{W_{+}}$ since it is the only tail index here.

Proposition \ref{prop evt} implies that 
\begin{equation}
\frac{\mathbf{Z}_{+}-b_{n}}{a_{n}}\overset{d}{\rightarrow }\mathbf{V}_{+},
\label{EVT1}
\end{equation}%
where $\mathbf{V}_{+}$ is jointly EV distributed with tail index $\xi $, and the
constants $a_{n}$ and $b_{n}$ are chosen in the proof of Proposition \ref%
{prop evt}.

Let $I=(I_{1},\ldots ,I_{k})\in \{1,\ldots ,T\}^{k}$ be the $k$ random
indices such that $Z_{n:n-j+1}=Z_{I_{j}}$, $j=1,\ldots ,k$, and let $\hat{I}$
be the corresponding indices such that $\hat{Z}_{n:n-j+1}=\hat{Z}_{\hat{I}%
_{j}}$. Then the convergence of $\mathbf{\hat{Z}}_{+}$ follows from (\ref{EVT1})
once we establish $|\hat{Z}_{\hat{I}_{j}}-Z_{I_{j}}|=o_{p}(a_{n})$ for $%
j=1,\ldots ,k$. We consider $k=1$ for simplicity and the argument for a
general $k$ is very similar. Denote $\varepsilon _{i}\equiv \hat{Z}%
_{i}-Z_{i} $.

Consider the case with $\xi >0$. the part in Assumption 2(v) for $\xi
\left. >\right. 0$ yields that%
\begin{eqnarray*}
\sup_{i}\left\vert \varepsilon _{i}\right\vert &=&\sup_{i}\left\vert \mathbf{%
X}_{i}\left( \beta _{0}-\hat{\beta}\right) \right\vert \\
&\leq &\left\vert \left\vert \beta _{0}-\hat{\beta}\right\vert \right\vert
\sup_{i}\left\vert \left\vert \mathbf{X}_{i}\right\vert \right\vert \\
&=&o_{p}(1).
\end{eqnarray*}%
Given this, we have that, on the one hand, $\hat{Z}_{\hat{I}}\left. =\right.
\max_{i}\{Z_{i}\left. +\right. \varepsilon _{i}\}\left. \leq \right.
Z_{I}\left. +\right. \sup_{i}\left\vert \varepsilon _{i}\right\vert \left.
=\right. Z_{I}\left. +\right. o_{p}(1)$; and on the other hand, $\hat{Z}_{%
\hat{I}}\left. =\right. \max_{i}\{Z_{i}\left. +\right. \varepsilon
_{i}\}\left. \geq \right. \max_{i}\{Z_{i}+\min_{i}\{\varepsilon
_{i}\}\}\left. \geq \right. Z_{I}+\min_{i}\{\varepsilon _{i}\}\left. \geq
\right. Z_{I}-\sup_{i}\left\vert \varepsilon _{i}\right\vert \left. =\right.
Z_{I}\left. -\right. o_{p}(1)$. Therefore, $|\hat{Z}_{\hat{I}}-Z_{I}|\leq
o_{p}(1)=o_{p}(a_{n})$ since $a_{n}\rightarrow \infty $.

Consider the case with $\xi =0$. Corollary 1.2.4 in de Haan and Ferreira (2007) implies
that $a_{n}=f_{W}\left( Q_{W}(1-1/n)\right) $. Thus, the part in Condition
2.3 for $\xi \left. =\right. 0$ implies that%
\begin{eqnarray*}
\frac{1}{a_{n}}\sup_{i}\left\vert \varepsilon _{i}\right\vert &\leq &\frac{%
\sup_{i}\left\vert \left\vert \mathbf{X}_{i}\right\vert \right\vert \cdot
\left\vert \left\vert \beta _{0}-\hat{\beta}\right\vert \right\vert }{%
f_{W}\left( Q_{W}(1-1/n)\right) } \\
&=&o_{p}(1).
\end{eqnarray*}%
Then the same argument as above yields that $\left\vert \hat{Z}_{\hat{I}%
}-Z_{I}\right\vert \left. \leq \right. O_{p}\left( \sup_{i}\left\vert
\varepsilon _{i}\right\vert \right) \left. =\right. o_{p}(a_{n})$. $%
\blacksquare $

\subsubsection*{Proof of Proposition \protect\ref{prop sym}}

Let $\mathbf{Z}_{-}^{\ast }$ denote the $k$ smallest order statistics of $%
\{Z_{i}\}$. Let $\left( a_{n}^{+},b_{n}^{+}\right) ^{\intercal }$ and $%
\left( a_{n}^{-},b_{n}^{-}\right) ^{\intercal }$ be the sequences of
normalizing constants for the right and left tails of $Z$, respectively.
Then by the same argument as in Proposition \ref{prop evtx}, we have $%
\mathbf{\hat{Z}}_{-}-\mathbf{Z}_{-}=o_{p}(a_{n}^{-})$ and $\mathbf{\hat{Z}}%
_{+}-\mathbf{Z}_{+}=o_{p}(a_{n}^{+})$. Therefore, it suffices to establish $%
\mathbf{Z}_{+}$ and $\mathbf{Z}_{-}$ jointly converge to $\left( \mathbf{V}%
_{+}^{\intercal },\mathbf{V}_{-}^{\intercal }\right) ^{\intercal }$ where $%
\mathbf{V}_{+}^{\intercal }$ and $\mathbf{V}_{-}^{\intercal }$ are
independent and both EV distributed with indices $\xi _{Z_{+}}$ and $\xi
_{Z_{-}}$, respectively. To this end, note that the case with $k=1$ is
established as Theorem 8.4.3 in Arnold et al. (1992). We now generalize their
argument for $k\geq 2$.

By elementary calculation and the i.i.d. assumption, the joint density of the
order statistics $Z_{n:n},\ldots ,Z_{n:1}$ is $n!\prod_{i=1}^{n}f_{Z}%
\left( z_{i}\right) $ for $z_{1}\leq z_{2}\leq \ldots \leq z_{n}$. Then by a
change of variables, the joint density of $(Z_{n:n}-b_{n}^{+})/a_{n}^{+},\ldots
,(Z_{n:n-k+1}-b_{n}^{+})/a_{n}^{+},(Z_{n:k}-b_{n}^{-})/a_{n}^{-},\ldots
,(Z_{n:1}-b_{n}^{+})/a_{n}^{+}$ satisfies that for $v_{1}^{-}\leq
v_{2}^{-}\leq \ldots \leq v_{k}^{-}\leq v_{k}^{+}\leq \ldots v_{1}^{+},$

\begin{eqnarray*}
&&\mathbb{P}\left( 
\begin{array}{l}
Z_{n:n}\leq a_{n}^{+}v_{1}^{+}+b_{n}^{+},...,Z_{n:n-k+1}\leq
a_{n}^{+}v_{k}^{+}+b_{n}^{+}, \\ 
Z_{n:1}\geq a_{n}^{-}v_{1}^{-}+b_{n}^{-},...,Z_{n:k}\leq
a_{n}^{-}v_{k}^{-}+b_{n}^{-}%
\end{array}%
\right) \\
&=&\left(
F_{Z}(a_{n}^{+}v_{k}^{+}+b_{n}^{+})-F_{Z}(a_{n}^{-}v_{k}^{-}+b_{n}^{-})%
\right) ^{n-2k} \\
&&\times \prod_{r=1}^{k}\left( n-r+1\right) a_{n}^{-}f_{Z}\left(
a_{n}^{-}v_{r}^{-}+b_{n}^{-}\right) \\
&&\times \prod_{r=1}^{k}\left( n-r+1\right) a_{n}^{+}f_{Z}\left(
a_{n}^{+}v_{r}^{+}+b_{n}^{+}\right) \\
&\equiv &P_{1n}\times P_{2n}\times P_{3n}\text{.}
\end{eqnarray*}

By the DOA assumption for both the left and right tails and equations
(8.3.1) and (8.4.9) in Arnold et al. (1992), 
\begin{equation*}
P_{1n}\rightarrow G_{\xi _{Z_{+}}}\left( v_{k}^{+}\right) \left( 1-G_{\xi
_{Z_{-}}}\left( v_{k}^{-}\right) \right) \text{.}
\end{equation*}%
By (8.4.4) in Arnold et al. (1992) and the fact that $k$ is fixed, $%
P_{2n}\rightarrow \prod_{r=1}^{k}g_{\xi _{Z_{-}}}\left( v_{r}^{-}\right)
/G_{\xi _{Z_{-}}}(v_{r}^{-})$ and $P_{3n}\rightarrow \prod_{r=1}^{k}g_{\xi
_{Z_{+}}}\left( v_{r}^{+}\right) /\left( 1-G_{\xi
_{Z_{+}}}(v_{r}^{+})\right) $. The proof is then complete by combining $%
P_{jn}$ for $j=1,2,3$ and the continuous mapping theorem. $\blacksquare $

\section{Computational details}

This section provides more details for constructing the test (\ref{test eq}), which is based on the limiting observations $\mathbf{V}_{-}^\ast$ and $\mathbf{V}_{+}^\ast$. The density is given by (\ref{density_xs}), which is computed by Gaussian Quadrature. To construct the test (\ref{test eq}), we specify the weight $w$ to be uniform over the alternative space for expositional simplicity, which can be easily changed. Then, it remains to determine a suitable candidate for the weight $\Lambda$ and the critical value cv$(k,\alpha)$. We do this by the generic algorithm provided by Elliott et al. (2015) and M\"{u}ller and Wang (2017). 

The idea of identifying a suitable choice of $\Lambda$ and cv$(k,\alpha)$ is as follows. First, we can discretize $\Xi $ into a grid $\Xi _{a}$ and determine $\Lambda$ accordingly as the point masses. Then we can simulate $N$ random draws of $\mathbf{V}_{-}^\ast$ and $\mathbf{V}_{+}^\ast$ from $\xi \in \Xi _{a}$ and estimate the rejection probability $P_{\xi }(\varphi_{\pm}
(\mathbf{V}_{-}^{\ast },\mathbf{V}_{+}^{\ast })=1)$ by sample fractions. The subscript $\xi$ emphasizes that the rejection probability depends on the value of $\xi$ that generates the data. By iteratively increasing or decreasing the point masses as a function of whether the estimated $P_{\xi }(\varphi_{\pm}
(\mathbf{V}_{-}^{\ast },\mathbf{V}_{+}^{\ast })=1)$ is larger or smaller than the nominal level, we can always find a candidate $\Lambda$ together with cv$(k,\alpha)$ that numerically satisfy the uniform size control.

In practice, we can determine the point masses by the following steps. Let $c$ be short for cv$(k,\alpha)$.
\\
{\bf Algorithm:}
\begin{enumerate}
\item Simulate $N=$ 10,000 i.i.d. random draws from some proposal density with $\xi $ drawn uniformly from $\Xi _{a}$, which is an equally spaced grid on $[0,0.99]$ with 50 points. 

\item Start with $\Lambda_{(0)}=\{1/50,1/50,\ldots ,1/50\}$ and $c=1$. Calculate the (estimated) coverage probabilities $P_{\xi_{j}}(\varphi _{\pm}(\mathbf{V}_{-}^{\ast },\mathbf{V}_{+}^{\ast })=1)$ for every $\xi _{j}\in \Xi _{a}$ using importance sampling. Denote them by $P=(P_{1},...,P_{50})^{\intercal }.$ 

\item Update $\Lambda $ and $c$ by setting $c\Lambda _{(s+1)}=c\Lambda _{(s)}+\kappa(P-0.05)$ with some step-length constant $\kappa>0$, so that the $j$-th point mass in $\Lambda $ is increased/decreased if the coverage probability for $\xi _{j}$ is larger/smaller than the nominal level. 

\item Integrate for 500 times. Then, the resulting $\Lambda_{(500)}$ and $c$ are a valid candidate. 

\item Numerically check if $\varphi _{\pm}$ with $\Lambda_{(500)}$ and $c$  indeed controls the size uniformly by simulating the rejection probabilities over a much finer grid on $\Xi $. If not, go back to step 2 with a finer $\Xi _{a}$. 
\end{enumerate}

\newpage

\renewcommand{\arraystretch}{1.2}

\begin{table}[tbp]
	\begin{center}
	
		\begin{tabular}{lcccccccc}
			\hline\hline
			$n$ & \multicolumn{3}{c}{100} &  &  & \multicolumn{3}{c}{1000} \\ 
			\cline{2-4}\cline{7-9}
			$k$ & 10 & 20 & 50 &  &  & 10 & 20 & 50 \\ \hline
			$F_{W}$ & \multicolumn{8}{c}{Rejection Prob. under half-normal $U_{i}$} \\ 
			N(0,1) & 0.01 & 0.00 & 0.00 &  &  & 0.03 & 0.02 & 0.01 \\ 
			La(0,1) & 0.05 & 0.04 & 0.02 &  &  & 0.05 & 0.05 & 0.05 \\ 
			t(2) & 0.31 & 0.45 & 0.35 &  &  & 0.30 & 0.49 & 0.76 \\ 
			$\pm $Pa(0.5) & 0.31 & 0.52 & 0.12 &  &  & 0.30 & 0.48 & 0.78 \\ 
			$\pm $F(4,4) & 0.32 & 0.50 & 0.50 &  &  & 0.29 & 0.50 & 0.80 \\ \hline
			$F_{W}$ & \multicolumn{8}{c}{Rejection Prob. under half-Laplace $U_{i}$} \\ 
			\hline
			N(0,1) & 0.02 & 0.00 & 0.00 &  &  & 0.03 & 0.01 & 0.01 \\ 
			La(0,1) & 0.05 & 0.04 & 0.050 &  &  & 0.06 & 0.06 & 0.04 \\ 
			t(2) & 0.32 & 0.46 & 0.29 &  &  & 0.32 & 0.50 & 0.80 \\ 
			$\pm $Pa(0.5) & 0.31 & 0.52 & 0.06 &  &  & 0.30 & 0.49 & 0.79 \\ 
			$\pm $F(4,4) & 0.31 & 0.52 & 0.49 &  &  & 0.28 & 0.48 & 0.78 \\ \hline
		\end{tabular}
	\end{center}		
\renewcommand{\arraystretch}{1}%
		
		\begin{singlespacing}%
			\caption{Small sample rejection probabilities of test (%
				\ref{test}) when there is no covariate. $U_{i}$ is generated from half-standard normal or
				half-Laplace(0,1) and $W_{i}$ is generated from standard normal,
				Laplace(0,1), Student-t(2), Pareto(0.5) and F(4,4). Based on 1000 simulation
				draws. Significance level is 0.05.}%
		\end{singlespacing}%

\end{table}

\renewcommand{\arraystretch}{1.2}%

\begin{table}[tbp]
	\begin{center}

		\begin{tabular}{lcccccccc}
			\hline\hline
			$n$ & \multicolumn{3}{c}{100} &  &  & \multicolumn{3}{c}{1000} \\ 
			\cline{2-4}\cline{7-9}
			$k$ & 10 & 20 & 50 &  &  & 10 & 20 & 50 \\ \hline
			$F_{W}$ & \multicolumn{8}{c}{Rejection Prob. under half-normal $U_{i}$} \\ 
			N(0,1) & 0.01 & 0.00 & 0.00 &  &  & 0.03 & 0.02 & 0.01 \\ 
			La(0,1) & 0.05 & 0.04 & 0.02 &  &  & 0.05 & 0.05 & 0.05 \\ 
			t(2) & 0.30 & 0.44 & 0.33 &  &  & 0.30 & 0.49 & 0.80 \\ 
			$\pm $Pa(0.5) & 0.32 & 0.51 & 0.10 &  &  & 0.30 & 0.48 & 0.80 \\ 
			$\pm $F(4,4) & 0.32 & 0.50 & 0.47 &  &  & 0.29 & 0.50 & 0.80 \\ \hline
			$F_{W}$ & \multicolumn{8}{c}{Rejection Prob. under half-Laplace $U_{i}$} \\ 
			\hline
			N(0,1) & 0.01 & 0.00 & 0.00 &  &  & 0.03 & 0.01 & 0.01 \\ 
			La(0,1) & 0.04 & 0.05 & 0.01 &  &  & 0.06 & 0.06 & 0.04 \\ 
			t(2) & 0.32 & 0.46 & 0.28 &  &  & 0.31 & 0.50 & 0.80 \\ 
			$\pm $Pa(0.5) & 0.32 & 0.50 & 0.07 &  &  & 0.30 & 0.49 & 0.80 \\ 
			$\pm $F(4,4) & 0.31 & 0.51 & 0.47 &  &  & 0.28 & 0.48 & 0.78 \\ \hline
		\end{tabular}
	\end{center}		
\renewcommand{\arraystretch}{1}

		\begin{singlespacing}%
			\caption{Small sample rejection probabilities of test (%
				\ref{test}) when there are covariates. $U_{i}$ is generated from half-normal and $W_{i}$
		                      is generated from standard normal, Laplace(0,1), Student-t(2),
				Pareto(0.5) and F(4,4). Based on 1000 simulation draws. Significance
				level is 0.05.}%
		\end{singlespacing}%

\end{table}

\renewcommand{\arraystretch}{1.2}%

\begin{table}[tbp]
	\begin{center}%
		
		\begin{tabular}{lccccccccc}
			\hline\hline
			$n$ &  & \multicolumn{3}{c}{100} &  &  & \multicolumn{3}{c}{1000} \\ 
			\cline{3-5}\cline{8-10}
			$k$ &  & 10 & 20 & 50 &  &  & 10 & 20 & 50 \\ \hline
			$F_{W}$ & \multicolumn{1}{l}{$F_{U}$} & \multicolumn{8}{c}{Rejection Prob.
				under $H_{0}$} \\ \hline
			N(0,1) & \multicolumn{1}{l}{half-N(0,1)} & 0.06 & 0.06 & 0.03 &  &  & 0.05 & 
			0.05 & 0.05 \\ 
			La(0,1) & \multicolumn{1}{l}{half-La(0,1)} & 0.04 & 0.05 & 0.07 &  &  & 0.05
			& 0.04 & 0.04 \\ 
			t(2) & \multicolumn{1}{l}{half-t(2)} & 0.04 & 0.06 & 0.24 &  &  & 0.06 & 0.06
			& 0.05 \\ 
			$\pm $Pa(0.5) & \multicolumn{1}{l}{$-$Pa(0.5)} & 0.05 & 0.06 & 0.67 &  &  & 
			0.05 & 0.06 & 0.05 \\ 
			$\pm $F(4,4) & \multicolumn{1}{l}{$-$F(4,4)} & 0.04 & 0.05 & 0.24 &  &  & 
			0.04 & 0.06 & 0.03 \\ \hline
			$F_{W}$ & \multicolumn{1}{l}{$F_{U}$} & \multicolumn{8}{c}{Rejection Prob.
				under $H_{1}$} \\ \hline
			N(0,1) & \multicolumn{1}{l}{$-$Pa(0.75)} & 0.28 & 0.68 & 0.99 &  &  & 0.25 & 
			0.55 & 0.94 \\ 
			Laplace(0,1) & \multicolumn{1}{l}{$-$Pa(0.75)} & 0.25 & 0.46 & 0.89 &  &  & 
			0.21 & 0.44 & 0.82 \\ 
			t(2) & \multicolumn{1}{l}{$-$Pa(0.75)} & 0.13 & 0.21 & 0.62 &  &  & 0.09 & 
			0.09 & 0.19 \\ 
			$\pm $Pa(0.5) & \multicolumn{1}{l}{$-$Pa(0.75)} & 0.10 & 0.23 & 0.87 &  &  & 
			0.07 & 0.12 & 0.17 \\ 
			$\pm $F(4,4) & \multicolumn{1}{l}{$-$Pa(0.75)} & 0.09 & 0.14 & 0.46 &  &  & 
			0.07 & 0.09 & 0.15 \\ \hline
		\end{tabular}
\end{center}		
\renewcommand{\arraystretch}{1}%

		\begin{singlespacing}%
			\caption{Small sample rejection probabilities of test (\ref{test eq}). $U_{i}$ is generated from half-norma, half-Laplace, Student-t(2), Pareto(0.5), F(4,4), and Pareto(0.75) and $W_{i}$
		                      is generated from standard normal, Laplace(0,1), Student-t(2),
				Pareto(0.5) and F(4,4).  Based on 1000 simulation draws. Significance level
				is 0.05.}%
		\end{singlespacing}%

\end{table}%

\renewcommand{\arraystretch}{1.2}%

\begin{table}[tbp]
	\begin{center}%	
	
		\begin{tabular}{lcccccccc}
			\hline\hline
			$n$ & \multicolumn{3}{c}{100} &  &  & \multicolumn{3}{c}{1000} \\ 
			\cline{2-4}\cline{7-9}
			$k$ & 10 & 20 & 50 &  &  & 10 & 20 & 50 \\ \hline
			$F_{U}$ & \multicolumn{8}{c}{Rejection Prob. under Normal $W_{i}$} \\ 
			half-N(0,1) & 0.02 & 0.01 & 0.00 &  &  & 0.02 & 0.02 & 0.00 \\ 
			half-La(0,1) & 0.05 & 0.03 & 0.00 &  &  & 0.06 & 0.05 & 0.04 \\ 
			half-t(2) & 0.31 & 0.45 & 0.46 &  &  & 0.34 & 0.54 & 0.88 \\ 
			$-$Pa(0.5) & 0.34 & 0.52 & 0.48 &  &  & 0.33 & 0.56 & 0.89 \\ 
			$-$F(4,4) & 0.31 & 0.52 & 0.70 &  &  & 0.32 & 0.52 & 0.86 \\ \hline
			$F_{U}$ & \multicolumn{8}{c}{Rejection Prob. under Laplace $W_{i}$} \\ \hline
			half-N(0,1) & 0.05 & 0.03 & 0.01 &  &  & 0.05 & 0.05 & 0.05 \\ 
			half-La(0,1) & 0.04 & 0.03 & 0.00 &  &  & 0.05 & 0.04 & 0.03 \\ 
			half-t(2) & 0.28 & 0.37 & 0.38 &  &  & 0.35 & 0.58 & 0.89 \\ 
			$-$Pa(0.5) & 0.30 & 0.40 & 0.44 &  &  & 0.33 & 0.58 & 0.89 \\ 
			$-$F(4,4) & 0.30 & 0.49 & 0.59 &  &  & 0.33 & 0.52 & 0.86 \\ \hline
		\end{tabular}		
	\end{center}		
\renewcommand{\arraystretch}{1}%
		
		\begin{singlespacing}%
			%EndExpansion
			\caption{Rejection probabilities of test (%
				\ref{test}). $U_{i}$ is generated from various distributions and $%
			W_{i}$  is generated from standard normal or\ Laplace(0,1). Based on
				1000 simulation draws. Significance level is 0.05.}%
		\end{singlespacing}%
		
\end{table}%

\renewcommand{\arraystretch}{1.2}%

\begin{table}[tbp]
	\begin{center}%
		
		\label{table:bank}
		
		\begin{tabular}{ccccccccccc}
			\hline\hline
			&  & \multicolumn{4}{c}{left tail} &  & \multicolumn{4}{c}{right tail} \\ 
			year & $k=$ & 25 & 50 & 75 & 100 &  & 25 & 50 & 75 & 100 \\ \hline
			1998 &  & $>0.1$ & 0.03 & 0.00 & 0.00 &  & 0.00 & 0.00 & 0.00 & 
			0.00 \\ 
			1999 &  & 0.00 & 0.00 & 0.00 & 0.00 &  & 0.05 & 0.00 & 0.00 & 0.00 \\ 
			2000 &  & $>0.1$ & 0.03 & 0.00 & 0.00 &  & 0.00 & 0.00 & 0.00 & 
			0.00 \\ 
			2001 &  & 0.00 & 0.00 & 0.00 & 0.00 &  & 0.00 & 0.00 & 0.00 & 0.00 \\ 
			2002 &  & $>0.1$ & 0.00 & 0.00 & 0.00 &  & $>0.1$ & 
			0.00 & 0.00 & 0.00 \\ 
			2003 &  & $>0.1$ & 0.05 & 0.00 & 0.00 &  & $>0.1$ & 
			0.00 & 0.00 & 0.00 \\ 
			2004 &  & 0.04 & 0.00 & 0.00 & 0.00 &  & $>0.1$ & 0.00 & 0.00 & 
			0.00 \\ 
			2005 &  & 0.09 & 0.00 & 0.00 & 0.00 &  & $>0.1$ & 0.00 & 0.00 & 
			0.00 \\ \hline
		\end{tabular}
			
		\renewcommand{\arraystretch}{1}%	
		\begin{singlespacing}%
			%EndExpansion
			\caption{P-values of the test (\ref{test}) for the US Banks data collected by Feng and Serletis (2009). }%	
		\end{singlespacing}%
	\end{center}		
\end{table}%

\end{document}